\documentstyle[twocolumn,prl,aps,epsf,epsfig,amsmath]{revtex}  

\font\cmss=cmss12   

\def\1{\hbox{{1}\kern-.25em\hbox{l}}}  

\def\bfZ{\relax{\hbox{\cmss Z\kern-.4em Z}}}  

\newcommand \Pomeron {I\!\!P}

\newcommand{\beq}{\begin{equation}}  

\newcommand{\eeq}{\end{equation}}  

\newcommand{\bea}{\begin{eqnarray}}   
\newcommand{\eea}{\end{eqnarray}}   
  
\begin{document}   
                  
\title{A next-to-leading  
order QCD analysis of deeply virtual Compton scattering amplitudes}   
 
\author{Andreas Freund\thanks{andreas.freund@physik.uni-regensburg.de}} 
\address{Institut f\"ur Theoretische Physik,  
University of Regensburg, Universit\"atstr. 31, 93053 Regensburg  
, Germany}   
\author{Martin McDermott\thanks{martinmc@amtp.liv.ac.uk}}  
\address{Division of Theoretical Physics, Dept. Math. Sciences, University of Liverpool, Liverpool, L69 3BX, UK}     
%\date{\today}   
\maketitle   
\begin{abstract}   
We present a next-to-leading order (NLO) QCD analysis of unpolarized and polarized deeply virtual Compton scattering (DVCS) amplitudes, for two different input 
scenarios, in the $\overline{MS}$ scheme. We illustrate and discuss  
the size of the NLO effects and the behavior of the amplitudes in skewedness,  
$\zeta$, and photon virtuality, $Q^2$.  
In the unpolarized case, at fixed $Q^2$, we find a remarkable effective 
power-law behaviour in $\zeta$, akin to Regge factorization, over several 
orders of magnitude in $\zeta$. We also quantify the ratio of real to imaginary parts 
of the DVCS amplitudes and their sensitivity to changes of the factorization scale.  
\vspace{1pc}   
\end{abstract}     
  
\section{Introduction}  
 
Deeply virtual Compton scattering (DVCS) \cite{mul94,rad,ji,diehl,van,fact,jios,ffs,exp}, 
$\gamma^* (q) + p (P) \rightarrow \gamma (q') + p (P') $, is the 
most promising process for accessing generalized parton distributions 
(GPDs) \cite{mul94,rad,ji,bmns,pet,ffgs} which carry new information about the   
dynamical degrees of freedom inside a nucleon. GPDs are an extension 
of the well-known parton distribution functions (PDFs) appearing  
in inclusive processes and are defined as the Fourier transform of   
{\it non-local} light-cone operators sandwiched between nucleon states of   
{\it different} momenta, commensurate with a finite momentum transfer 
in the t-channel. These distributions are true two-particle 
correlation functions and contain, in addition to the usual PDF-type  
information residing in the DGLAP \cite{dglap} region,   
supplementary information about the distribution amplitudes of virtual 
meson-like states in the nucleon in the ERBL \cite{erbl} region.  

We recently presented a full numerical 
solution of the associated renormalization group equations at next-to-leading  
order (NLO) accuracy, for unpolarized and polarized distributions,  
using realistic input models \cite{frmc1}, as well as a complete NLO 
QCD analysis of DVCS observables \cite{frmc2}. To achieve this we  
calculated the real and imaginary parts of unpolarized and polarized DVCS 
amplitudes, ${\cal T}^{V/A}_{DVCS}$, at NLO which are related to the  
triple differential cross section on the lepton level 
(which also includes the Bethe-Heitler (BH) process) via 
\begin{align}  
&\frac{d\sigma^{(3)} (e^{\pm} p \rightarrow e^{\pm} \gamma p)}{dx_{bj} dQ^2d|t|} 
= \int^{2\pi}_0d\phi \frac{d\sigma^{(4)}}{dx_{bj} dQ^2d|t|d\phi} \nonumber\\ 
&=\frac{\alpha_{e.m.}^3 x_{bj} y^2}{8\pi Q^4}\left(1+\frac{4M^2 x_{bj}^2}{Q^2}\right)^{-1/2} \int^{2\pi}_0d\phi|{\cal T}^{\pm}|^2 \, ,   
\label{crossx}  
\end{align} 
\noindent where  
\bea  
&|{\cal T}^{\pm}|^2 = \nonumber  
&|{\cal T}_{DVCS}|^2 \pm ({\cal T}^*_{DVCS}{\cal T}_{BH} + {\cal T}_{DVCS} {\cal T}^*_{BH}) + |{\cal T}_{BH}|^2 \, ,    
\label{tdef}  
\eea  
\noindent and $M$ is the proton mass.
The dependent variables $Q^2 = - q^2, x_{bj} = Q^2/2 P \cdot q, t$ and $\phi$ are minus the photon 
virtuality, Bjorken $x$, the momentum transfer to the proton squared and the 
relative angle between the lepton and proton scattering planes \cite{bemu2}, 
respectively.  Variable $y = Q^2/x_{bj} S$, where $S$ is the total lepton-proton center of mass energy. 
For the DVCS process in the proton rest frame, $y$ is the fraction of the energy of the 
incoming lepton carried by the photon.
 
The real and imaginary parts of the polarized and unpolarized DVCS amplitudes  
are interesting in their own right, since each can be independently accessed 
experimentally via various asymmetries which exploit the $\phi$-dependence of the 
interference term \cite{ffs,bemu3}. Thus the predictions that we give here for 
the real and imaginary parts of the amplitudes at NLO, their behavior in the skewedness, 
$\zeta=x_{bj}$, and $Q^2$, as well  
as the relationship between real and imaginary parts, which all depend on the GPDs,  
can be directly tested by experiment. A NLO analysis of the DVCS amplitudes for  
large $\zeta$ was carried out in \cite{bemu1} and, at the limited points where a comparison is possible, we agree with their results. 
 
This paper is structured as follows. In Section \ref{sec2} we define the input model  
for the GPDs and briefly review their NLO evolution. We state   
the necessary factorized convolution integrals and explain their  
exact technical implementation, via subtractions. In Section \ref{sec3} we give detailed NLO results for the real and imaginary parts of the unpolarized (subsection \ref{sec3a}) and polarized (subsection \ref{sec3b}) DVCS amplitudes, comparing them to LO results using the same input GPDs and commenting on their sensitivity to the input GPDs and the factorization scale, $\mu^2$. We also give the $Q^2$ and $\zeta$ dependence of the ratio of  
real to imaginary parts and quantise the importance of the ERBL region to the real part. We close our discussion in Section \ref{sec4} with a statement on the general structure of  
NLO and NNLO corrections and briefly conclude in Section \ref{conc}.  
For convenience in an appendix  
(section \ref{secapp}) we restate the NLO coefficient functions \cite{bemu1} (in \ref{secapp1}) and give analytic results for their integrals (in \ref{secapp2}), which are required to implement the subtractions in section \ref{sec3}.  
   
\section{Factorization Theorem and Definitions} 

\label{sec2} 

There are many representations for GPDs in the literature \cite{mul94,rad,ji,bmns,pet,ffgs,golbier}.  
We chose to work in a particular representation identical to the non-diagonal 
representation defined in \cite{golbier} which is a natural one when 
comparing to experiments. We use 
\begin{align}  
&{\cal F}^{S(a),V/A} (X,\zeta,\mu^2,t) = \nonumber\\ 
&\left[\frac{H^{a,V/A} (v,\zeta,\mu^2,t) \mp H^{a,V/A} (-v,\zeta,\mu^2,t)}{(1-\zeta/2)}\right] \, , \nonumber \\  
&{\cal F}^{g,V/A} (X,\zeta,\mu^2,t) = \nonumber\\ 
&\left[\frac{H^{g.V/A} (v,\zeta,\mu^2,t) \pm H^{g,V/A} (-v,\zeta,\mu^2,t)}{(1-\zeta/2)}\right] \, ,   
\label{defsinglet}  
\end{align} 
where $v = (X -\zeta/2)/(1-\zeta/2)$, $S(a)$ and $g$ 
refer to the quark singlet for flavour $a$ and the gluon, 
respectively, $V$ and $A$ stand for unpolarized  
(vector) and polarized (axial-vector) cases, taking the upper and lower signs,   
respectively. This representation is different from the usual one  
(see for example \cite{bmns}) which is defined symmetrically with respect to 
the incoming and outgoing nucleon plus momentum (defined on the interval  
$v \in [-1,1]$ and symmetric about $v=0$). The GPDs in eq.~(\ref{defsinglet}) 
have plus momentum fractions (on the interval $X\in [0,1]$) with respect to  
the incoming nucleon momentum, $P$, in analogy to the PDFs of inclusive  
reactions, with the ERBL region in the interval $X \in [0,\zeta]$ and 
the DGLAP region in the interval $X \in [\zeta,1]$. The transformation 
between the symmetric and non-diagonal representation is given in \cite{golbier}.  
Furthermore, within the non-diagonal representation  
$\zeta = x_{bj} = -q^2/ 2P \cdot q $
and the symmetry of the GPDs, which was previously manifest about $v=0$, is  
now manifest about the point $X=\zeta/2$.  
 
We build input distributions, ${\cal F}^{S(a), g} (X,\zeta, Q_0^2)$, at the  
input scale, $Q_0$, with the correct symmetries and properties, from conventional  
PDFs in the DGLAP region, for both the unpolarized and polarized cases,  
by employing the factorized model due to Radyushkin \cite{radmod},  
which is based on double distributions. The $H$ functions (symmetric GPDs)  
required for eq.(\ref{defsinglet})  
are related to the latter via the following reduction formula (factoring out  
the overall $t$-dependence):   
\begin{align}   
H (v, \zeta) = \int^{1}_{-1} dx' \int^{1-|x'|}_{-1+|x'|} dy' \delta \left( x' + \frac{\zeta y'}{2-\zeta} - v \right) F (x',y') \, . 
\label{reduction}  
\end{align}   
   
The double distributions, $F^{i,V/A}$, are a product of a profile function, $\pi^{i}$, and a conventional PDF, $f^{i,V/A}$, $(i = q(a), g$):   
\bea  
F^{q(a)}(x',y') &=& \pi^{q} (x',y') f^{q(a)} (x') \nonumber\\   
    &=& \frac{3}{4} \frac{(1-|x'|)^2 - {y'}^2}{(1-|x'|)^3} f^{q(a)} (x') \, , \nonumber\\   
F^g(x',y')  &=& \pi^g(x',y') f^g (x') \nonumber\\   
    &=& \frac{15}{16} \frac{((1-|x'|)^2 - {y'}^2)^2}{(1-|x'|)^5} f^g (x') \, , \label{DDs}  
\eea  
\noindent where   
\bea   
f^g (x)  &=& xg(x,Q_0) \Theta(x) + |x| g (|x|,Q_0) \Theta (-x) \nonumber \\   
f^{q(a)} (x) &=& q^{a} (x,Q_0) \Theta(x) - (\bar q^{a}) (|x|,Q_0) \Theta (-x) \, . \label{pdfinp}  
\eea   
The profile functions are chosen to guarantee the correct symmetry properties   
in the ERBL region which are preserved under evolution, as we explicitly illustrated in \cite{frmc1}. Their normalization is specified by demanding that the   
conventional distributions are reproduced in the forward limit at the input scale:      
${\cal F}^{i} (X,\zeta \rightarrow 0, Q_0) \rightarrow f^i (X,Q_0)$. 
 
In addition to the contributions from the double distributions the  
unpolarized singlet GPDs also contain a so-called ``D-term''  
\cite{vanderhagen1,poly}, which is only non-zero in the ERBL region, and  
ensures the correct polynomiality in $\zeta$ \cite{poly} of the moments in 
$X$ of the GPDs. There is an equivalent term in the unpolarized gluon distribution 
but apart from its symmetry nothing is known about this function, thus we 
chose to set it to zero. 
 
Within the above class of input model, we specify two particular input 
models for the GPDs by using two sets of inclusive 
unpolarized/polarized PDFs (for use in eqs.(\ref{defsinglet},\ref{reduction},\ref{DDs})), i.e. GRV98/GRSV00 \cite{grv} with $\Lambda^{(4,NLO)}_{{\mbox{\tiny QCD}}} = 246~\mbox{MeV}$ and MRSA'/GS(A) \cite{mrsap} with $\Lambda^{(4,NLO)}_{{\mbox{\tiny QCD}}} = 231~\mbox{MeV}$ at the common input scale  
$Q^2_0 = 4~\mbox{GeV}^2$ and $\Lambda^{(4,LO)}_{{\mbox{\tiny QCD}}} =
174~\mbox{MeV}$ for both sets. Using two different choices allows us to 
investigate the sensitivity of the amplitudes to the choice of 
input. The GPDs are then evolved in LO and NLO using our newly 
developed evolution code \cite{frmc1}. Note that there are two sets of  
evolution equations which have to be solved simultaneously, one for the 
ERBL region \cite{erbl} and one for the DGLAP \cite{dglap} region,  
with the ERBL one being dependent on the  
evolution in the DGLAP region, whereas the evolution in the DGLAP region  
is independent from the evolution in the ERBL region (for more details  
on the NLO skewed evolution and NLO coefficient functions see \cite{frmc1,nlos}). 
 
The factorization theorem \cite{fact,jios} proves that the DVCS amplitude takes the following factorized form (in the non-diagonal representation) up to terms suppressed by ${\cal O} (1/Q)$:  
\begin{align}  
&{\cal T}^{S,V/A}_{DVCS} (\zeta,Q^2,\mu^2,t) = \sum_a e^2_a \left(\frac{2 - \zeta}{\zeta} \right) \Big[ \Big.  \nonumber\\ 
&\int^1_0 dX~T^{S(a),V/A} \left(\frac{2X}{\zeta} - 1, \frac{Q^2}{\mu^2} \right) ~{\cal F}^{S(a),V/A} (X,\zeta,\mu^2,t) \mp \nonumber\\  
&\Big. \int^1_{\zeta} dX~T^{S(a),V/A} \left(1 - \frac{2X}{\zeta},\frac{Q^2}{\mu^2}\right)~{\cal F}^{S(a),V/A} (X,\zeta,\mu^2,t) \Big] ,\nonumber\\  
&{\cal T}^{g,V/A}_{DVCS} (\zeta,Q^2,\mu^2,t) = \frac{1}{N_f}\left (\frac{2 - \zeta}{\zeta}\right )^2  \Big[ \Big. \nonumber\\ 
&\int^1_0 dX~T^{g,V/A} \left(\frac{2X}{\zeta} - 1, \frac{Q^2}{\mu^2} \right) ~{\cal F}^{g,V/A} (X,\zeta,\mu^2,t) \pm \nonumber\\ 
&\Big. \int^1_{\zeta} dX~T^{g,V/A} \left(1 - \frac{2X}{\zeta}, \frac{Q^2}{\mu^2}\right) ~{\cal F}^{g,V/A}(X,\zeta,\mu^2,t) \Big] \, .  
\label{tdvcs}  
\end{align} 
 
In the following we will initially set the factorization scale, $\mu^2$, equal  
to the photon virtuality, $Q^2$, and will later investigate its variation.  
Henceforth, we will suppress the factorized $t$-dependence since all of our predictions will be made 
for $t=0$. The LO and NLO coefficient functions, $T^{i,V/A}$, are  
taken from eqs.(14-17) of \cite{bemu1} and are summarized in the appendix \ref{secapp1}.  
They can have both real and imaginary parts, depending on the region of integration, which in turn generates 
real and imaginary parts of the DVCS amplitudes. For the integrals over the range $X \in [0,1]$,
the coefficient functions contain singularities associated with the point $X=\zeta$,  which are regulated using a ``$+ i \epsilon$'' prescription. For our numerical integration, we choose to regulate 
these integrals using the Cauchy principal value prescription (denoted $P.V.$) as follows: 
\begin{align} 
&\Big. P.V. \int^1_0 dX~T\left(\frac{2X}{\zeta} - 1\right) {\cal F}(X,\zeta,Q^2) = \nonumber\\  
&\int^{\zeta}_0 dX~T\left(\frac{2X}{\zeta} - 1\right)\left({\cal 
F}(X,\zeta,Q^2)-{\cal F}(\zeta,\zeta,Q^2)\right) + \nonumber\\ 
&\int^1_{\zeta} 
dX~T\left(\frac{2X}{\zeta} - 1\right) \left({\cal F}(X,\zeta,Q^2) -{\cal F}(\zeta,\zeta,Q^2)\right) + \nonumber \\ 
&{\cal F}(\zeta,\zeta,Q^2)\int^1_0 dX~T \left(\frac{2X}{\zeta} - 1\right) \, . 
\label{subtraction} 
\end{align} 
Each term in eq.~(\ref{subtraction}) is now either separately finite 
or only contains an integrable logarithmic singularity.  
This algorithm closely resembles the implementation of the $+$ regularization in the  
evolution of PDFs and GPDs. Note that the first integral in eq.~(\ref{subtraction})  
(in the ERBL region) is strictly real, however, the second and third terms contain both real and imaginary parts (which are generated in the DGLAP region).  
This definition leads to the following formulas for the real and 
imaginary parts of the DVCS amplitudes: 
 
\begin{align} 
&\mbox{Re}~{\cal T}^{S,V/A}_{DVCS} (\zeta,Q^2)  = \sum_a e^2_a ~\left( \frac{2 - \zeta}{\zeta} \right) \Bigg[ \nonumber\\ 
&\int^{\zeta}_0 dX ~T^{S,V/A}\left(z\right)\left({\cal F}^{S,V/A} (X,\zeta)-{\cal F}^{S,V/A} (\zeta,\zeta) \right)+ \nonumber\\ 
&\int^1_{\zeta}dX\Big[\mbox{Re}T^{S,V/A} \left(z\right)\left({\cal
F}^{S,V/A} (X,\zeta)-{\cal F}^{S,V/A} (\zeta,\zeta) \right)\nonumber\\
&\mp T^{S,V/A}\left(-z\right) {\cal F}^{S,V/A} (X,\zeta)\Big] \nonumber\\ 
&+ {\cal F}^{a,V/A} (\zeta,\zeta) ~\mbox{Re}~\int^1_0 dX~T^{S,V/A} \left(z\right) \Bigg]  \, , \nonumber\\  
&\mbox{Im}~{\cal T}^{S,V/A}_{DVCS} (\zeta,Q^2) = \sum_a e^2_a ~\left( \frac{2 - \zeta}{\zeta} \right) \Bigg[ \nonumber\\ 
&\int^1_{\zeta}dX\Big[\mbox{Im}T^{S,V/A} \left(z\right)\left({\cal
F}^{S,V/A} (X,\zeta)-{\cal F}^{S,V/A} (\zeta,\zeta) \right) \Big] \nonumber\\
&+ {\cal F}^{S,V/A} (\zeta,\zeta) ~\mbox{Im} \int_{0}^{1} dX T^{S,V/A} \left(z\right)  \Bigg] \, , \label{reimampqs} 
\end{align} 
\begin{align} 
&\mbox{Re}~{\cal T}^{g,V/A}_{DVCS} (\zeta,Q^2) = \frac{1}{N_f} \left(\frac{2 - \zeta}{\zeta}\right )^2 \Bigg[  \nonumber\\ 
&\int^{\zeta}_0 dX ~T^{g,V/A}\left(z\right) \left({\cal F}^{g,V/A} 
(X,\zeta)-{\cal F}^{g,V/A}(\zeta,\zeta)\right)+\nonumber\\ 
&\int^1_{\zeta}dX\Big[\mbox{Re}T^{g,V/A} \left(z\right)\left({\cal
F}^{g,V/A} (X,\zeta)-{\cal F}^{g,V/A} (\zeta,\zeta) \right)\nonumber\\
&\pm T^{g,V/A}\left(-z\right) {\cal F}^{g,V/A} (X,\zeta)\Big] \nonumber\\ 
&+ {\cal F}^{g,V/A}(\zeta,\zeta) ~\mbox{Re}\int^1_0 dX~T^{g,V/A}\left(z\right) \Bigg] \, , \nonumber\\ 
&\mbox{Im}~{\cal T}^{g,V/A}_{DVCS} (\zeta,Q^2) = \frac{1}{N_f} \left(\frac{2 - \zeta}{\zeta}\right )^2  \Bigg[  \nonumber\\ 
&\int^1_{\zeta}dX\Big[\mbox{Im}T^{g,V/A} \left(z\right)\left({\cal
F}^{g,V/A} (X,\zeta)-{\cal F}^{g,V/A} (\zeta,\zeta) \right) \Big] \nonumber\\ 
&+ {\cal F}^{g,V/A}(\zeta,\zeta) ~\mbox{Im} \int^1_0 dX~T^{g,V/A} \left(z \right) \Bigg] \, , 
\label{reimampg} 
\end{align} 
\noindent where $z=2X/\zeta -1$ and for convenience we have suppressed the scale and explicit quark flavour dependence of the GPDs on the right hand sides. 
The real and imaginary parts of the  
unpolarized and polarized DVCS 
amplitudes were computed using a FORTRAN code based on numerical  
integration routines (for more details see \cite{frmcbig}). We 
implemented the exact solution to the RGE equation for $\alpha_s$, in 
LO and NLO as appropriate, in our calculation to be consistent throughout our analysis. 
 
\section{NLO and LO DVCS amplitudes} 
\label{sec3} 
 
In the next two subsections we present results for the real and imaginary 
parts of the NLO DVCS amplitudes for the unpolarized and polarized cases, 
comparing them to LO results (using the same input GPDs) throughout. 
This allows us to quantify the effect of moving from LO to NLO.
We first plot the absolute values, then the ratio of real to imaginary 
parts (in $Q^2$ for fixed $\zeta$ and in $\zeta$ for fixed $Q^2$). 
Finally, we discuss the influence of the ERBL region on the real part of the  
amplitudes and the factorization scale dependence.  
 
\subsection{The unpolarized DVCS amplitudes} 
\label{sec3a} 
 
\begin{figure} 
\centering 
\mbox{\epsfig{file=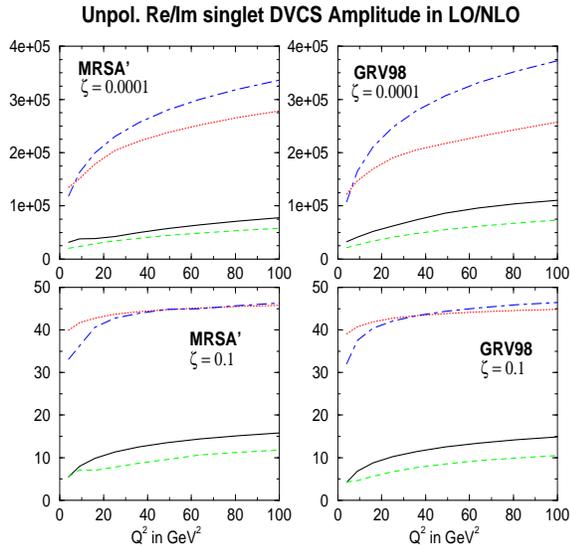,width=7.5cm,height=7.5cm}} 
\caption{The $Q^2$-dependence of the real and imaginary parts of the quark singlet DVCS 
amplitude. The solid (dashed) curve is the real part in LO (NLO) and 
the dotted (dashed-dotted) curve is the imaginary part in LO (NLO).} 
\label{ampunpolsq} 
\end{figure}

In Fig.(\ref{ampunpolsq}) we plot the real and imaginary parts of the  
unpolarized quark singlet amplitude (cf. eq.~(\ref{reimampqs})) at LO and NLO. 
We observe that MRSA' and GRV98 give broadly similar results.

The NLO corrections are generally fairly large and tend to decrease the value of the real part and increase the value of the imaginary part of the quark singlet amplitude (bearing in mind that the NLO imaginary amplitude is lower at the input scale, due to the inclusion of the NLO coefficient function).
The relative correction, $R$, in moving from LO to NLO (i.e. $R =$ (NLO-LO)/LO)) can be inferred from Fig.(\ref{ampunpolsq}). $R$ is typically $-20\%$ to $-40 \%$ for the real part for both values of $\zeta$. For the imaginary part at $\zeta = 0.1$ the corrections are moderate ($|R| < 20 \%$), whereas at small $\zeta$, $R$ can be as large as $+50 \%$ for GRV98 at large $Q^2$.
The amplitudes drop dramatically in going from small to large $\zeta$ reflecting the strong decrease in the GPDs. They generally increase with increasing $Q^2$, albeit moderately, reflecting the expected $\ln(Q^2)$ behavior. In fact, an approximate scaling is observed  at LO and NLO, but only sets in at large $Q^2$ for small $\zeta$ in the imaginary part at NLO. 

In Fig.(\ref{ampunpolsq}) and subsequent figures a comparison of the LO and NLO curves at the input scale $Q^2=4$~GeV$^2$ 
reveals the effect of including NLO coefficient functions alone. Of course at higher $Q^2$ NLO effects 
are included in both the coefficient functions and in the evolution of the GPDs. For the numerical size of the NLO 
corrections in the evolution of the GPDs alone see \cite{frmc1}.
 
\begin{figure} 
\centering 
\mbox{\epsfig{file=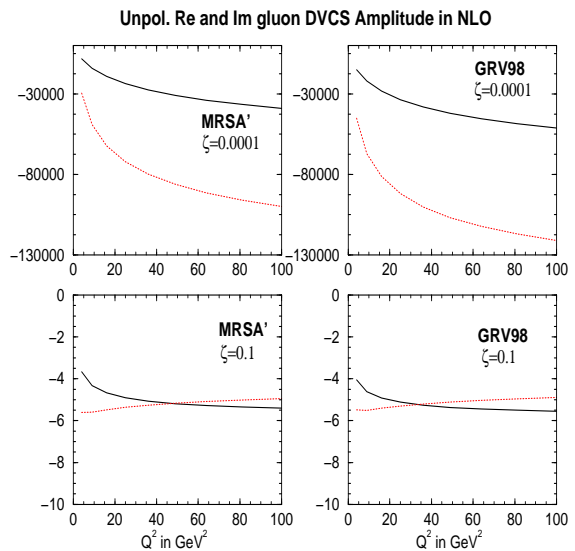,width=7.5cm,height=7.5cm}} 
\caption{The $Q^2$-dependence of the real (solid line) and imaginary (dotted line) parts of the unpolarized gluon DVCS 
amplitude, for two representative values of $\zeta$.} 
\label{ampunpolgq} 
\end{figure}

In Fig.(\ref{ampunpolgq}) we show the real and imaginary parts of the  
unpolarized gluon amplitude (cf. eq.~(\ref{reimampg})) which starts at NLO. 
We note firstly that the gluon contribution is of the same order of magnitude as the quark singlet one, although it is suppressed by $\alpha_s/2\pi$ and secondly that both the real and imaginary parts are large and negative. This explains the strong variation of the azimuthal angle asymmetry (AAA), in moving from LO to NLO, observed in \cite{frmc2} (for the case of a dominant Bethe-Heitler contribution) which is directly proportional to the real part of a combination of DVCS amplitudes. Otherwise, the gluon mirrors the NLO quark singlet amplitudes in its behavior in $Q^2$ and for large and small $\zeta$. 
 
Of course for the physical amplitude at NLO one must add the quark singlet and gluon 
contributions together. For the real part at small $\zeta=10^{-4}$ this leads to a relative 
relative change in moving from LO to NLO of $40\%$ (MRSA') and $80\%$ (GRV98) at the input scale 
to $75\%$ (MRSA') and $80\%$ (GRV98) at $Q^2=100$~GeV$^2$. For large $\zeta= 0.1$ the relative 
change in the real part at both scales is about $60\%$.  
For the imaginary part at small $\zeta= 10^{-4}$ the relative changes for MRSA' are $35\%$ 
(at the input) and $15\%$ (at $Q^2=100$~GeV$^2$). For GRV98 one finds $50\%$ (input scale) and 
$2\%$ (evolved scale) changes. For the imaginary part at large $\zeta=0.1$, for both MRSA' and GRV98, one finds about $30\%$ (input scale) and $10\%$ (evolved scale) changes.
Note that the NLO changes in the physical amplitude decrease as $Q^2$ increases in line 
with the expectation of perturbative QCD that the NLO corrections should die out as  
$Q^2\rightarrow \infty$. 

\begin{figure} 
\centering 
\mbox{\epsfig{file=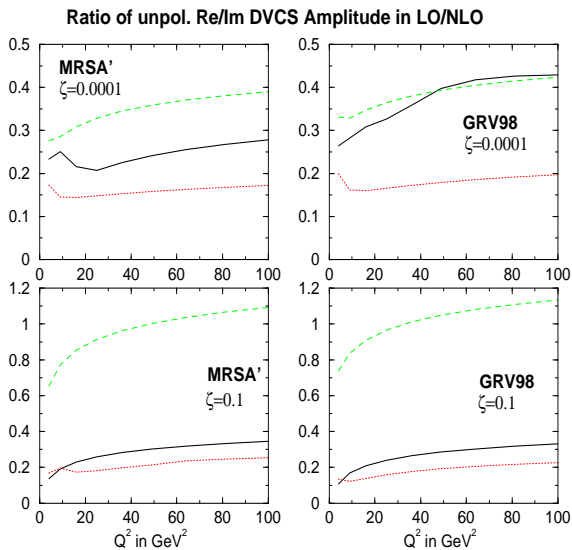,width=7.5cm,height=7.5cm}} 
\caption{The ratio of real to imaginary parts of the unpolarized quark 
singlet and gluon DVCS amplitudes, as a function of $Q^2$. The solid (dotted) curve is 
for the quark singlet in LO (NLO) and the dashed curve is for the gluon in NLO.} 
\label{reimunpolq} 
\end{figure}

In Fig.~(\ref{reimunpolq}) we show the ratio of the real to imaginary parts for  
both the quark singlet and the gluon amplitudes as a function of $Q^2$. 
We note that for small $\zeta$ the ratios can be as large as $45\%$ 
(a similar value was found for the closely related process of high energy 
$J/\psi$ photoproduction in \cite{fms}). 
The quark ratios are slightly different for the two inputs.  
The greatest contrast is seen at small $\zeta$:  
the NLO case is basically flat in $Q^2$ which differs markedly from LO which rises with $Q^2$. Both cases are similar at large $\zeta$. The gluon ratio is remarkably  
similar for the two input sets, at both small and large $\zeta$,  
for the whole $Q^2$ range considered. 

\begin{figure} 
\centering 
\mbox{\epsfig{file=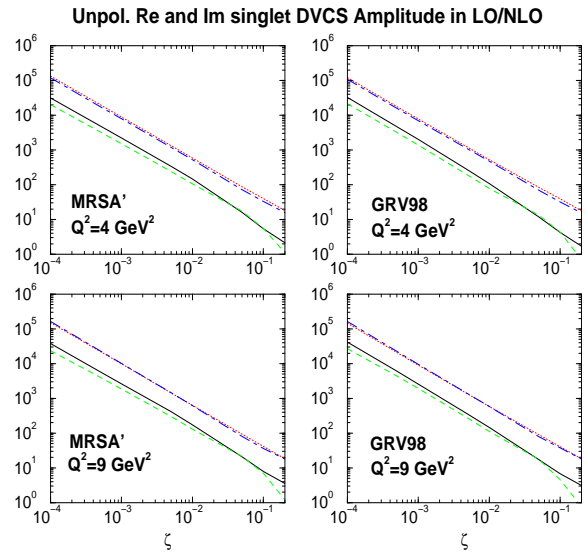,width=7.5cm,height=7.5cm}} 
\caption{The real and imaginary parts of the unpolarized  quark singlet DVCS 
amplitude, as a function of $\zeta$. The solid (dashed) curve is the real part in LO 
(NLO) and the dotted (dashed-dotted) curve is the imaginary part in LO (NLO).  
A remarkably simple behaviour is observed in $\zeta$ which is close to a single power over a wide range. Because $Q^2$ is fixed, this behaviour in $\zeta$ translates directly into a single power in energy.} 
\label{ampunpolsx} 
\end{figure}

\begin{figure} 
\centering 
\mbox{\epsfig{file=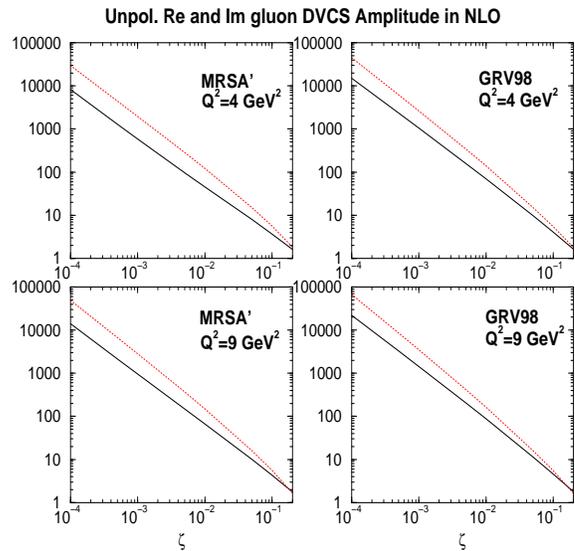,width=7.5cm,height=7.5cm}} 
\caption{The modulus of the real and imaginary parts of the unpolarized gluon DVCS amplitude as  functions of $\zeta$, for fixed $Q^2$.  
The solid curve is the modulus of the real part and the dotted curve is the modulus of the imaginary part of the gluon amplitude.} 
\label{ampunpolgx} 
\end{figure}

We now turn to the $\zeta$-dependence for fixed $Q^2$ which is shown in   
fig.(\ref{ampunpolsx}) for the quark singlet and fig.(\ref{ampunpolgx}) for the gluon. 
The most striking feature for the quark singlet case is that the amplitudes exhibit 
an effective power-like behavior in $\zeta$ over basically the {\it whole} range ($\zeta \in [0.0001,0.2]$), as illustrated  by the straight lines in fig.(\ref{ampunpolsx}).  
A simple two parameter fit of the type $a_0\zeta^{\lambda_0}$ works remarkably well  
up to about $\zeta=0.1$. The best fit is obtained with a four parameter fit, of the 
type $c_1 \zeta^{\lambda_1} (1 + c_2 \zeta^{\lambda_2})$, which can reproduce most of the  
curves on the few percent level, with $\lambda_0$ and $\lambda_1$ within $5\%-20\%$ of each other, and $\lambda_2$ small. The simple two parameter  
fit works best for the imaginary part of the amplitudes where we  
obtain a value for $\lambda_0$ between $-1.1$ and $-1.25$ with a  
moderate growth in $Q^2$ as expected from measurements of the slope 
of $F_2$ and HERA diffractive processes. For the real part of the amplitude the 
simple fit starts to decrease in quality around $\zeta=0.05$ (depending on the 
input). 

Similar power-like behaviors have been observed in many small-$x$ processes at  at HERA, including diffractive DIS in which the $x_{\Pomeron}$-dependence of diffractive structure functions is known to factorize for small $x_{\Pomeron}$. This behavior, known as Regge factorization, is predicted for high energy processes within Regge theory given the postulate of a universally exchanged Regge pole, known as the Pomeron. Since the relationship between the phenomenological Regge Theory and perturbative QCD  
remains unresolved, we find the observation of this single power in our
numerical perturbative QCD calculation remarkable and very interesting.

Note that we do not claim to have derived this powerlike behaviour from first principles. 
The analytic forms for the coefficients functions (see appendix \ref{secapp1}) would seem to favor a more complicated sum of logs in $\zeta$, particularly when the interplay with the effective 
behaviour in $\zeta$ of the GPDs, at a given $Q^2$, is taken into account.
Hence, the fact that a single power apparently works for such a large range in $\zeta$ for 
DVCS is somewhat surprising. Naively one may expect that a sum of two or more powers would 
be required to give a reasonable fit. Perhaps this indicates that DVCS always proceeds  
through partonic configurations which lie in the same universality class,   
i.e. are self-similar. The reason why these self-similar configuration seem to be  
of importance beyond the ``diffractive region'' can be understood if one examines  
the behavior of the integrand in the convolution integrals in  
eqs.~(\ref{reimampqs}, \ref{reimampg}). One observes that a large contribution to 
the integral and ultimately to the imaginary part of the amplitude itself stems  
from the region in $X$ very close to $\zeta$, even for larger values of 
$\zeta$, leading to the self-similar behavior.  
This is not too surprising given the steep rise of the GPD in the 
DGLAP region towards $\zeta$ and that the coefficient 
functions are singular at $\zeta$, where even the implementation of the 
principal value in Sec.\ (\ref{sec2}) leaves an integrable 
singularity. 

This situation is somewhat altered for the real part of the 
amplitude where the ERBL region plays a very important part, as we 
will see. There, although the coefficient function is singular, 
the symmetry of the GPD in the ERBL region makes the value of the 
integral somewhat less dependent on the region near $\zeta$, i.e. less 
singular, especially for large $\zeta$. Although the value of the real part also 
depends on an integral over the DGLAP region which contributes more to  
the power like behavior, this dependence progressively decreases as 
$\zeta$ grows. 
 
The physical picture which is emerging is the following: the 
region near $\zeta$ corresponds to large light-cone distances for the 
operators which translates into two parton configurations, either a 
$q\bar q$ pair in the ERBL region or a quark leaving and then returning to 
the proton in the DGLAP region, where one of the partons, either the
$\bar q$ or the returning $q$, carries 
virtually zero momentum fraction, $X-\zeta$, in either the $+$ or $-$ direction on  
the light cone and the other carries approximately $\zeta$, 
i.e. very asymmetric configurations seem to have a disproportionate  
weighting in the amplitude. 
The coefficient functions, with their 
singularity structure at $\zeta$, weight these configurations much more  
heavily in the amplitude than other, more symmetric, configurations.  

Although asymmetric configurations become rare as one approaches the valence region of  
large $\zeta$, because they should be mainly found in the sea, they are still enhanced by the 
singularity structure of the coefficient functions.  
This then leads us to the conclusion that, although the sea is small at large $\zeta$, DVCS  
still proceeds largely through sea configurations in the valence region,  
thus its relative scarcity compared to DIS in the valence region.  
Physically, at large $X \approx \zeta$, one has to strike an unusual fluctuation  
in the proton to emit a real photon, while still leaving the proton intact. 

For the gluon contribution, shown in Fig.(\ref{ampunpolgx}), we find the same behavior as in the case of the quark singlet. 
Note that we took the modulus of both the real and imaginary parts, since they are actually {\it negative}, in order to produce a log-log plot.  
Performing the same type of fits as in the quark case, one obtains similar numbers for $\lambda_0$ between $-1.14$ and $-1.28$ in the two parameter fit and again very similar ones in 
the four parameter fit ($5\%-25\%$ variation in the powers).  
The quality of the two parameter fit starts to decrease rapidly for a $\zeta \sim 0.05$. 
Nevertheless, the explanation given in the quark singlet case is still applicable in the case of the gluon. 

\begin{figure} 
\centering 
\mbox{\epsfig{file=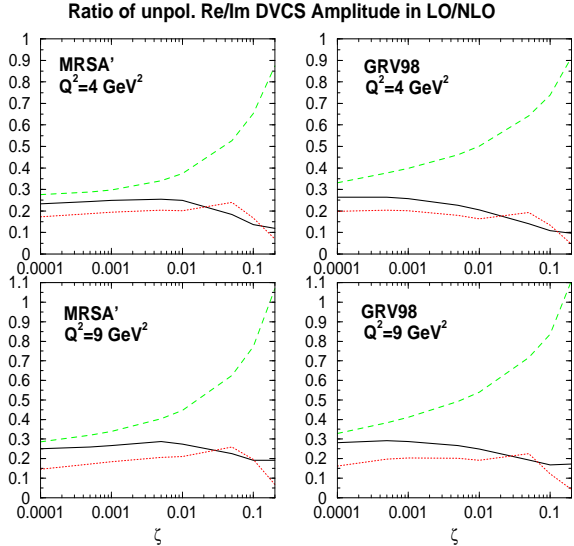,width=7.5cm,height=7.5cm}} 
\caption{The ratio of real to imaginary parts of the unpolarized quark 
singlet and gluon DVCS amplitude, as a function of $\zeta$, at fixed $Q^2$.  
The solid (dotted) curve is 
the ratio in LO (NLO) and the dashed curve is the ratio for the gluon in NLO.} 
\label{reimunpolx} 
\end{figure}

In Fig.~(\ref{reimunpolx}) we show the ratio of real to 
imaginary parts in $\zeta$ for fixed $Q^2$. Again we note the remarkable  
similarity between the gluon curves for both inputs, both in shape as well  
as absolute values. 
A very mild growth is seen for the quark singlet  
ratio in $\zeta$, whereas the gluon varies quite strongly in $\zeta$.  

Using Fig.~(\ref{reimunpolx}) we can make a simple test to check whether we 
have computed the real part of the amplitude at small $\zeta$ properly 
by employing a dispersion relation for the unpolarized amplitudes at  
small $\zeta$ (for more details see \cite{fms}): 
\begin{align} 
\frac{\mbox{Re}~T_{DVCS}}{\mbox{Im}~T_{DVCS}} = 
\tan\left(\frac{\pi}{2}(|\lambda|-1)\right) \, . 
\label{disrel} 
\end{align}   
For the fitted values of $\lambda$ we obtain a ratio which is in very 
good agreement with the values in Fig.~(\ref{reimunpolx}) up to about  
$\zeta = 5\times 10^{-3}$ for both the quark singlet and the gluon.  
This confirms the self-consistency of our calculation. 

Next we discuss the relative importance of the ERBL region to the value of 
the real part of the DVCS amplitude, starting with the quark singlet. 
On inspection of the relative contribution of the ERBL  
integrals ($X \in [0,\zeta]$) in eqs.~(\ref{reimampqs}, \ref{reimampg}) 
we find that at small $\zeta$ the ERBL region integral has a relative 
contribution  between $90\%$ at the input scale and $140\%$ at
$Q^2=100~\mbox{GeV}^2$ ($100\%$ and $50\%$ respectively in LO) of the
value of the amplitude, i.e. there is a large cancellation between 
the subtraction term and the $X \in [\zeta,1]$ integral with both of them 
being substantially larger, individually than the  $[0,\zeta]$ integral.  
As one increases $\zeta$ the relative importance of the ERBL region 
drops to $50\%$ at the input scale and $130\%$ at $Q^2=100~\mbox{GeV}^2$ 
($80\%$ and $30\%$ respectively in LO), however,  the subtraction term 
now starts to dominate the value of the amplitude. This observation is in line with our previous
argument of the importance of very asymmetric parton configurations. 
Remember, that the subtraction term is directly 
proportional to the GPD at $\zeta$. Also note that going from LO to NLO  
seems to change the relative importance of the ERBL region, especially  
its $Q^2$ behavior. 

Turning now to the gluon we make a 
slightly different observation. Firstly, at small $\zeta$ the relative  
contribution varies from $40\%-60\%$ in going from the input scale to $Q^2=100~\mbox{GeV}^2$ 
again there is a large cancellation between the previously mentioned 
terms, but the change in $Q^2$ is not very dramatic. Increasing 
$\zeta$ one finds an increase of the relative importance to  
$60\% - 80\%$ in going from the input scale to our large $Q^2$ value. 
Again the subtraction term becomes more and more dominant, relatively 
speaking, and thus our interpretation from the quark case carries over
to the gluon case. 
 
To close our analysis of the unpolarized case,  
we discuss the scale dependence of the DVCS amplitudes.  
We varied the factorization scale, $\mu^2$, from $Q^2$, used above, to $Q^2/2$
and $2Q^2$ for both sets and found the following variations, where the two 
sets agree fairly well with one another. 
At small $\zeta$, we found a small variation at the input scale of $3-5\%$ 
which increases to about $15\% - 30\%$ at large $Q^2$ (we used $Q^2=100~\mbox{GeV}^2$, for the large scale, throughout) for both real and imaginary part of the quark singlet. For the gluon we find a variation of 
about $15\%$ for the real and imaginary part at the input scale which reduces 
to about $10\%$ for the real part and to about $2\%$ for the imaginary part.   
At large $\zeta$, we find similar variations as the authors of \cite{bemu1}, 
i.e. around $5\%$ for both the real and imaginary parts of the quark singlet 
and around $10\%$ for the gluon, for both the input scale and at large $Q^2$. 
In summary one can say that the scale dependence is not troublesome and that 
the uncertainties due to the chosen GPD are still much larger than those due 
to the factorization scale.

\subsection{The polarized case} 
\label{sec3b} 
 
\begin{figure} 
\centering 
\mbox{\epsfig{file=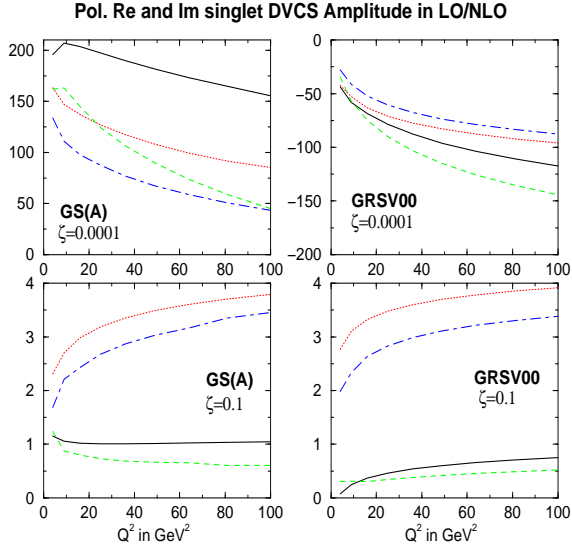,width=7.5cm,height=7.5cm}} 
\caption{Real and imaginary parts of the polarized quark singlet DVCS 
amplitude, as a function of $Q^2$, for fixed $\zeta$.  
The solid (dashed) curve is the real part in LO (NLO) and 
the dotted (dashed-dotted) curve is the imaginary part in LO (NLO).} 
\label{amppolsq} 
\end{figure}

For the polarized case we proceed exactly as we did for the 
unpolarized case starting in Fig.~(\ref{amppolsq}) with the absolute 
values of the real and imaginary parts of the polarized quark singlet  
amplitudes as functions of $Q^2$. Again the relative impact of the NLO 
corrections to the quark singlet can be inferred from this figure.
One immediately notices for small $\zeta$ the very large NLO corrections 
for the GS(A) input ($|R|$ can be as large as $70 \%$), whereas for 
the GRSV00 input (`standard scenario', i.e. unbroken sea) the corrections 
are more moderate ($|R| < 20 \%$). 
The very large corrections can be easily explained since in \cite{frmc1}  
it was shown that the GPD evolution drastically alters the shape of the  
GS(A) distribution in fact inverting its shape, whereas the shape of  
GRSV00 was almost unchanged and the absolute value changed only moderately.  
The difference in shape at small $\zeta$ was due to radically different  
assumptions about the polarized sea distribution.
At large $\zeta$, where the shape of the two input sets is similar 
we find that the shape of the amplitudes in $Q^2$ is also very similar, 
whereas their absolute values differ.  

\begin{figure} 
\centering 
\mbox{\epsfig{file=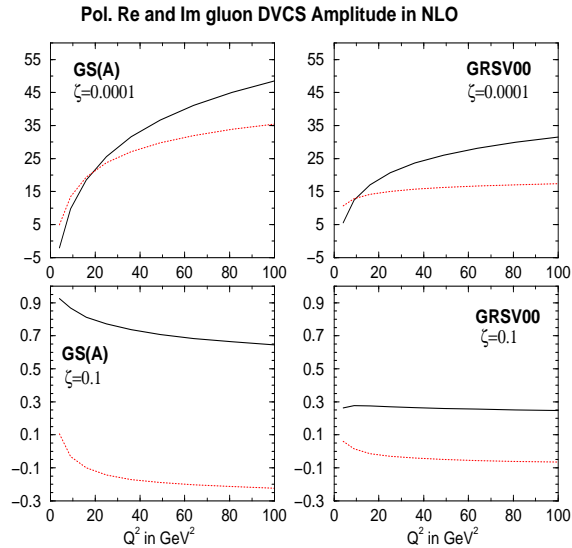,width=7.5cm,height=7.5cm}} 
\caption{The real and imaginary parts of the polarized gluon DVCS 
amplitude, as a function of $Q^2$, for fixed $\zeta$. The solid curves show the real part and the dotted curves show the imaginary part of the gluon amplitude.} 
\label{amppolgq} 
\end{figure}

The real and imaginary parts of the polarised gluon amplitude are plotted in Fig.~(\ref{amppolgq}).
Note that the real and imaginary parts of the amplitude are positive for small $\zeta$, exactly the opposite to the unpolarized case. For large $\zeta$, the real parts are positive, again in contrast to the unpolarised case. 
The imaginary part at large $\zeta$ starts positive but becomes negative 
under evolution. This behavior is due to the particular shape of the polarized gluon GPD at small and large $\zeta$ (see \cite{frmc1}). 

\begin{figure} 
\centering 
\mbox{\epsfig{file=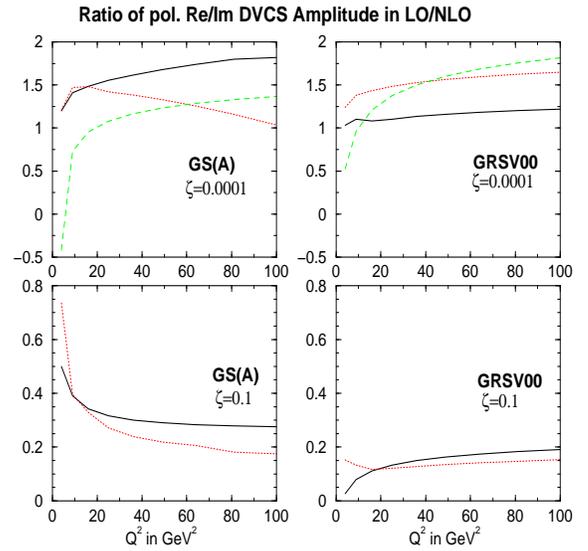,width=7.5cm,height=7.5cm}} 
\caption{The ratio of real to imaginary parts of the polarized quark 
singlet and gluon DVCS amplitudes, as a function of $Q^2$, at fixed $\zeta$.  
The solid (dotted) curves show the ratio for the quark singlet in LO (NLO) and the dashed curves show the ratio for the gluon in NLO.} 
\label{reimpolq} 
\end{figure}

In Fig.(\ref{reimpolq}) we show the ratio of real to imaginary parts  
as a function of $Q^2$, for fixed $\zeta$. 
At large $\zeta$, we again observe flattening at large  
$Q^2$: the GRSV00 result is flatter than  
the GS(A) case, especially at NLO, where we still observe strong  
variations in $Q^2$.  
For large $Q^2$, the gluon ratio is 
very similar for both sets and fairly flat in $Q^2$.
We did not plot the gluon ratio at large $\zeta$ since it has a 
very large value near the point where the imaginary part changes sign 
and thus would have completely swamped the quark result which we find 
more interesting here. Note that our LO ratio for the GS(A) model is 
in agreement with the values obtained in \cite{bemu3}.  

\begin{figure} 
\centering 
\mbox{\epsfig{file=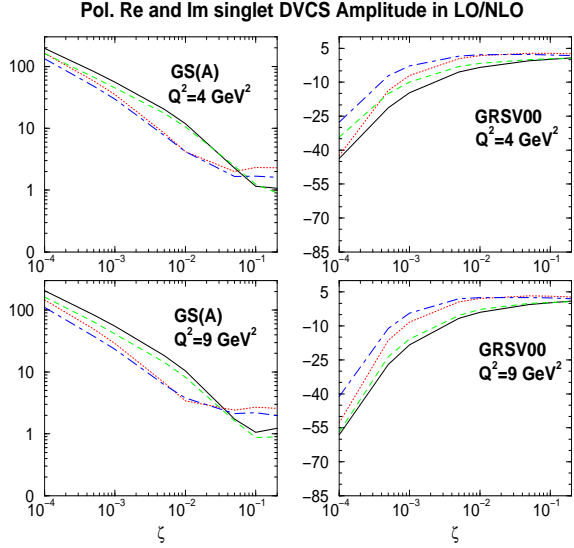,width=7.5cm,height=7.5cm}} 
\caption{Polarized real and imaginary parts of the quark singlet DVCS 
amplitude, as a function of $\zeta$, for fixed $Q^2$. The solid (dashed) curve is the real part in LO 
(NLO) and the dotted (dashed-dotted) curve is the imaginary part in LO (NLO).} 
\label{amppolsx} 
\end{figure}

In Fig.~(\ref{amppolsx}) we plot the $\zeta$-behaviour at fixed $Q^2$ and find a single power-like behaviour only for very small $\zeta$ (up to about $3\times 10^{-3}$), seen explicitly on the log-log plot for GS(A).  
Performing the same type of fits as in the unpolarized case, i.e. a simple two and four parameter fits, reveals a $\lambda_0$ between $-0.4$ to $-0.55$, 
and a similar story for the four parameter fits. 
However, the second power, $a_2$, is now substantially larger than in the 
unpolarized case, in order to be able to describe the large $\zeta$ 
behavior. The four parameter fit is able to describe the behavior of the  
amplitudes on the few percent level. Thus the simple single-power Regge-type effective behavior observed in the unpolarized case is only valid at small $\zeta$ in the polarized case. This is because the polarized sea dies out even more quickly with increasing $\zeta$, than the unpolarized one: therefore the 
highly asymmetric configurations necessary to produce a universal 
behavior in $\zeta$ become very rare.  
At large $\zeta$ they are almost completely gone.

\begin{figure} 
\centering 
\mbox{\epsfig{file=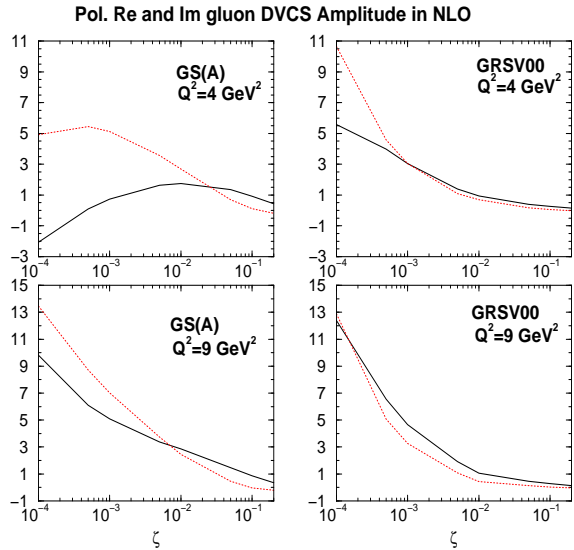,width=7.5cm,height=7.5cm}} 
\caption{Real and imaginary parts of the polarized gluon DVCS 
amplitude, as a function of  $\zeta$, at fixed $Q^2$.  
The solid curve is the real part in NLO and the 
dotted curve is the imaginary part.} 
\label{amppolgx} 
\end{figure}

Turning to the gluon in Fig.~(\ref{amppolgx}) we illustrate that the  
behavior in $\zeta$ at an evolved scale, $Q^2=9$~GeV$^2$, is very similar in shape and size for the two input sets, despite the fact that they start off very  different at the input scale, $Q^2=4$~GeV$^2$.   
The evolution forces the distributions to be quite similar very 
quickly.

For the GS(A) input, at small $\zeta$, the relative changes in going from LO 
to NLO in the physical polarized amplitude, i.e. the sum of polarized quark 
singlet and polarized gluon, are found to be about $15\%$ for both the real 
and imaginary parts at the input scale and about $8\%$ for the imaginary part 
and $40\%$ for the real part at $Q^2=100$~GeV$^2$. At large $\zeta=0.1$, the changes in the 
real part vary from about $90\%$ at the input scale to $25\%$ at $Q^2=100$~GeV$^2$ and those  
in the imaginary part vary from about $25\%$ (at the input scale) to $15\%$ at the evolved scale.
For GRSV00 we also find a decrease in the variation which at small $\zeta$ drops from $60\%$
at the input scale to $25\%$  at $Q^2=100$~GeV$^2$ for the imaginary part,  
and from $35\%$ to $5\%$ for the real part. At large $\zeta$ there is a 
decrease from about $25\%$, at the input for both real and imaginary parts, to about 
$2\%$ and $15\%$ for the real and imaginary parts, respectively, at the evolved scale. 
As in the unpolarized case the NLO corrections to the physical amplitude decrease as 
$Q^2$ increases except in the case of the real part at small $\zeta$ for GS(A) which is in 
line with the dramatic NLO evolution effects in the ERBL region. 

The relative change induced by NLO corrections of the quark singlet can be inferred directly 
from Fig. \ref{amppolsx}.

\begin{figure} 
\centering 
\mbox{\epsfig{file=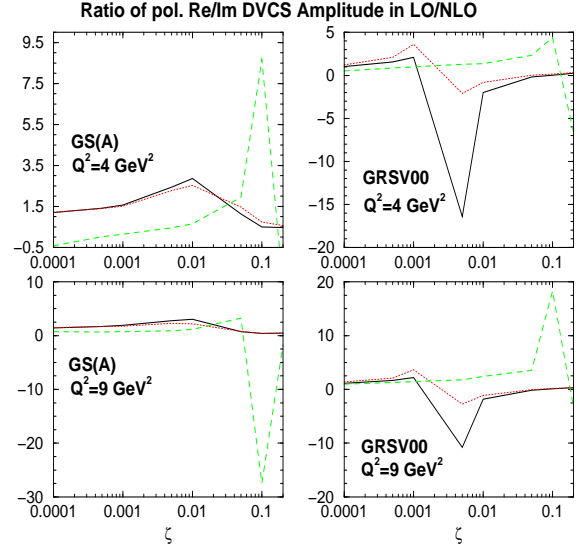,width=7.5cm,height=7.5cm}} 
\caption{The ratio of real to imaginary parts of the polarized quark 
singlet and gluon DVCS amplitudes, as a function of $\zeta$ at fixed $Q^2$. The solid (dotted) curve is the ratio in LO (NLO) and the dashed curve is the ratio for the gluon, which starts at NLO.} 
\label{reimpolx} 
\end{figure} 

The $\zeta$-dependence of the ratio of real to imaginary parts, plotted in 
Fig.~(\ref{reimpolx}), is rather different to the unpolarized case, with a strong fluctuation  
in the gluon due to a sign change of one of the amplitudes around the spike (we sample at a finite number of points in $\zeta$). The real part is typically much larger then the imaginary part, in contrast to the unpolarized case.  

Concerning the issue of the importance of the ERBL region, we find a  
similar picture to the unpolarized case, although the relative  
percentage contribution is smaller by about a factor of two, for both  
quarks and gluons. The influence of the subtraction term on the value  
of the real part of the amplitude is considerably less than in the  
unpolarized case, thus the ERBL region is weighted more heavily in  
the final value of the amplitude.  
This is in line with our observation of the deviations of the amplitudes  
from a single power in $\zeta$, for relatively small values of $\zeta$,  
compared to the unpolarized case (for which the ERBL region was of less 
importance as compared to the subtraction term). 
 
Finally, we comment briefly on the factorization scale dependence of the polarized amplitudes. 
We proceed in an identical manner to the unpolarized case in varying $\mu^2$. 
For small $\zeta$, we find a similar variation as in the unpolarized 
case for the quark singlet and larger variations of $35-40\%$ at the input
scale and $1-30\%$ at large $Q^2=100$~GeV$^2$, respectively, for the gluon. 
At large $\zeta$, we find variations in the real part of the quark singlet around 
$100\%$ at the input and at large $Q^2$. The imaginary part only varies 
between $5\%$ at the input scale to $20\%$ at the evolved scale. Both real and 
imaginary parts of the gluon amplitude at large $\zeta$ are very well behaved 
and vary only between $5-20\%$ at both the input scale and at large $Q^2$.
It is encouraging that the variations due to factorization scale changes  
can be safely neglected since the polarized distributions are even
less well known than the unpolarized ones. 
 
\section{General comments on NLO and NNLO corrections} 
\label{sec4} 
 
In this section we make some general comments about the 
structure of NLO corrections and expected NNLO corrections.  
 
In \cite{frmc1} we pointed out that the relative shape change in the 
NLO evolution of GPDs, compared with LO, is due to a new class of integrable 
divergences, $\ln\left(1-X/\zeta\right)^n/(1-X/\zeta)^i, 
~\mbox{n,i}=0,1,2$, appearing in the region around $\zeta$.  
The same type of integrable divergence also appears in the NLO  
coefficient functions, but is absent at LO, although one has an integrable  
singularity of the $1/(1-X/\zeta)$ type. This fact alone helps 
to explain why one finds, in certain regions, large changes in going from 
LO to NLO in the amplitudes.  
 
For DVCS observables the appearance of the gluon at NLO also changes  
things dramatically since the gluon contribution turns out to be of the same  
order as the quarks, at least at small $\zeta$ due to an extra factor of  
$1/\zeta$ in eq.~(\ref{tdvcs}).  
So, not only is a new quantity introduced, but one of the same  
order of magnitude as our LO quantities. 
Furthermore, with the unpolarized real part of the gluon amplitude being {\it 
negative} and of comparable size at small $\zeta$ as the NLO correction to the 
real part of the quark amplitudes (which is also negative), observables sensitive 
to the real part, such as the azimuthal angle asymmetry are expected to change 
dramatically in NLO (see \cite{frmc2}).  
Conversely, the effect should not be as dramatic for observables 
sensitive to the imaginary part, despite the fact that the NLO gluon correction
is big and {\it negative}, since the correction to the quark singlet at NLO is big and 
positive so the corrections will cancel to a certain extent (see \cite{frmc2}). 
 
What would one expect in NNLO ? From experience obtained 
in calculating forward coefficient functions, and some evolution 
kernels at ${\cal O}(\alpha^2_s)$, one would expect to see only the same type  
of integrable divergences reappearing, maybe with different powers in 
logs and rational functions, but not a new class of divergences which 
could radically alter the behavior of the amplitudes. Also, in contrast to moving from LO to NLO which gives the first gluon contributions, no new parton species appears at NNLO. This leads us to speculate that the NNLO order  
corrections should be mild. 
  
\section{Conclusions} 
\label{conc} 
 
We have presented a detailed analysis of the quark singlet and gluon  
contributions to the polarized and unpolarized DVCS amplitudes at NLO,  
using NLO-evolved generalised partons distributions (GPDs) built from sensible  
input models. We have compared throughout with the LO results using the same  
input GPDs, and have therefore quantized the effect of the NLO corrections. 
 
These results are directly relevant to measurable quantities in $ep \rightarrow ep\gamma$ processes at the HERA and HERMES experiments,  
and hence may be used to constrain the GPDs at NLO. 
 
The most striking feature of our results is that for a given $Q^2$ 
the unpolarized amplitudes exhibit an effective single-power behaviour in the 
skewedness parameter, $\zeta$,  
over a very large range, apparently indicating a universal behaviour. 
 
\section*{Acknowledgements}   
   
\label{sec:ack}   

A.~F. was supported by the DFG under contract $\#$ FR 1524/1-1. M.~M. was  
supported by PPARC. We are happy to acknowledge discussions with D.~Mueller 
and M.~Diehl.  
   
\section{Appendix} 
\label{secapp} 
 
\subsection{LO and NLO coefficient functions} 
\label{secapp1} 
 
The coefficient functions in eq.~(\ref{tdvcs}) are expanded in powers  
of $\alpha_s$. Up to terms of ${\cal O}(\alpha^2_s)$, they read
\begin{align} 
T^{q, V/A}(z) =& T^{q,LO}(z) + \frac{\alpha_s(\mu^2)}{2\pi}T^{q, NLO, V/A}(z,Q^2/\mu^2) \, , \nonumber\\  
T^{g, V/A}(z) =& T^{g,LO}(z) + \frac{\alpha_s(\mu^2)}{2\pi}T^{g, NLO, V/A}(z,Q^2/\mu^2) \, , 
\end{align} 
with $z = 2X/\zeta -1$. 
 
The LO and NLO coefficient functions were taken from \cite{bemu1} and 
are given by: 
\begin{align} 
&T^{q,LO} (z) = \frac{1}{1-z},\nonumber\\ 
&T^{g,LO} (z) = 0,\nonumber\\ 
&T^{q,NLO,V} (z) = T^{q,NLO,A} (z) - \frac{C_F}{1+z}\ln\frac{1-z}{2},\nonumber\\ 
&T^{q,NLO,A} (z) = \frac{C_F}{2(1-z)}\Big[\left(2\ln\frac{1-z}{2} + 
3\right) \times \nonumber\\ 
&\left(\ln\frac{Q^2}{\mu^2} + \frac{1}{2}\ln\frac{1-z}{2} -  
\frac{3}{4}\right) - \frac{27}{4} - \frac{1-z}{1+z}\ln\frac{1-z}{2}\Big] \, ,\nonumber\\ 
&T^{g,NLO,V} (z) = - T^{g,NLO,A} (z) + \nonumber\\  
&\frac{N_F}{2}\Big[\frac{1}{1-z}\left(\ln\frac{Q^2}{\mu^2} + 
\ln\frac{1-z}{2} - 2\right) + \frac{\ln\frac{1-z}{2}}{1+z}\Big] \, ,\nonumber\\ 
&T^{g,NLO,A} (z) = \frac{N_F}{2}\Big[\left(\frac{1}{1-z^2}  
+ \frac{\ln\frac{1-z}{2}}{(1+z)^2}\right) \times \nonumber\\  
&\left(\ln\frac{Q^2}{\mu^2} + \ln\frac{1-z}{2} - 2\right) - 
\frac{\ln^2\frac{1-z}{2}}{2(1+z)^2}\Big], 
\label{coef}  
\end{align} 
where the LO quark coefficient is normalized in such a way that, in the  
forward limit, after properly restoring the dependence on both 
skewedness parameters, one recovers the LO DIS coefficient 
$\delta(1-x)$ and the NLO gluon coefficient is normalized such that 
one recovers $\frac{1}{2}C^{DIS}_g$. 
 
In the interval $[0,\zeta]$, the above coefficients are strictly real. 
However, in the interval $[\zeta,1]$, they split into a real and imaginary parts,   
which can be easily deduced from eq.~(\ref{coef}).

\subsection{LO and NLO subtraction functions} 
\label{secapp2}   
 
In this section, we present the subtraction functions needed in our 
implementation of the Cauchy principal value prescription of eq.(\ref{subtraction}),  
i.e. the integrals 
\begin{align} 
&I^{q,V/A}(\zeta)= \int^1_0 dX~T^{q,V/A} \left(\frac{2X}{\zeta}-1\right) \, , \nonumber\\ 
&I^{g,V/A}(\zeta)= \int^1_0 dX~T^{g,V/A} \left(\frac{2X}{\zeta}-1\right) \, .  
\end{align}    
There are four different integrals to be done, which are regulated by adding 
$+ i \epsilon$ to $X$ (this implies analytically continuing the log terms in the {\it lower half plane} to pick up $-i \pi$). 
This generates real and imaginary parts for the subtraction functions, $I^{q,g}$, given below, for the unpolarised (V) and polarised (A) cases. 
One has to be careful to take the appropriate sheet of the Riemann surface 
for the logarithms in order to obtain the correct imaginary parts, i.e. to use the $+i\epsilon$ prescription consistently.  
In order to ensure the correctness of the results below, we cross checked them both with MATHEMATICA and MAPLE. 
\begin{align} 
&I^{q,LO}(\zeta) = -\frac{\zeta}{2} \Big[\ln\left(\frac{1-\zeta}{\zeta}\right) - i\pi \Big] \, , 
\end{align} 
\begin{align} 
&I^{q,NLO,A}(\zeta,Q^2,\mu^2) = \frac{C_F \zeta}{4} \Bigg[ \frac{\pi^2}{6} - \mbox{Li}_2\left(1-\frac{1}{\zeta}\right)  + \nonumber\\ 
& \ln\left(\frac{1-\zeta}{\zeta}\right) (\pi^2 + 9 + \ln\zeta - \frac{1}{3}\ln^2\left(\frac{1-\zeta}{\zeta}\right)) 
 + \nonumber\\ 
& \ln\frac{Q^2}{\mu^2}\left(\pi^2 - 3\ln\left(\frac{1-\zeta}{\zeta}\right) - \ln^2\left(\frac{1-\zeta}{\zeta}\right)\right) - \nonumber\\ 
&i\pi\Big[\frac{\pi^2}{3} + 9 + \ln\zeta - \ln^2\left(\frac{1-\zeta}{\zeta}\right) - \nonumber\\ 
& \quad \ln\frac{Q^2}{\mu^2}\left( 2 \ln\left(\frac{1-\zeta}{\zeta}\right) + 3 \right)\Big] \, \Bigg], 
\end{align} 
\begin{align} 
&I^{q,NLO,V}(\zeta,Q^2,\mu^2) = \frac{C_F \zeta}{4} \Bigg[\frac{\pi^2}{2} - 3\mbox{Li}_2\left(1-\frac{1}{\zeta}\right) + \nonumber\\ 
& \ln\left(\frac{1-\zeta}{\zeta}\right) (\pi^2 + 9 + 3 \ln\zeta - \frac{1}{3}\ln^2\left(\frac{1-\zeta}{\zeta}\right))        
 + \nonumber\\ 
&\ln\frac{Q^2}{\mu^2}\left(\pi^2 - 3\ln\left(\frac{1-\zeta}{\zeta}\right) - \ln^2\left(\frac{1-\zeta}{\zeta}\right)\right) - \nonumber\\ 
&i\pi\Big[\frac{\pi^2}{3} + 9 + 3 \ln\zeta - \ln^2\left(\frac{1-\zeta}{\zeta}\right) - \nonumber\\ 
& \quad \ln\frac{Q^2}{\mu^2}\left( 2 \ln\left(\frac{1-\zeta}{\zeta}\right) + 3 \right) \Big] \, \Bigg] \, , 
\end{align} 
\begin{align} 
&I^{g,NLO,A}(\zeta,Q^2,\mu^2) = \frac{N_F \zeta}{4} \Bigg[1 + \frac{\pi^2 \zeta}{4} + \nonumber\\ 
&\zeta \ln\left(\frac{1-\zeta}{\zeta}\right) (1 - \frac{1}{4}\ln \left(\frac{1-\zeta}{\zeta}\right) ) - \nonumber\\ 
&\frac{1}{2}\ln\frac{Q^2}{\mu^2}\left(1 + \zeta \ln\left(\frac{1-\zeta}{\zeta}\right)\right) - \nonumber\\ 
& \frac{i\pi\zeta}{2} \Big[2 - \ln\left(\frac{1-\zeta}{\zeta}\right) - \ln\frac{Q^2}{\mu^2} \Big] \, \Bigg], 
\end{align} 
\begin{align} 
&I^{g,NLO,V}(\zeta,Q^2,\mu^2) = \frac{N_F \zeta}{4} \Bigg[-1 + \frac{\pi^2}{3} \left(1 - \frac{3\zeta}{4} \right) + \nonumber\\ 
&\mbox{Li}_2\left(1-\frac{1}{\zeta}\right) - \ln\zeta \ln\left(\frac{1-\zeta}{\zeta}\right) +\nonumber\\ 
&(2-\zeta) \ln\left(\frac{1-\zeta}{\zeta}\right) (1 - \frac{1}{4}\ln\left(\frac{1-\zeta}{\zeta}\right) ) + \nonumber\\ 
& \frac{1}{2}\ln\frac{Q^2}{\mu^2}\left(1 - (2-\zeta)\ln\left(\frac{1-\zeta}{\zeta}\right)\right) - \nonumber\\ 
&\frac{i\pi}{2} \Big[(2-\zeta)\left(2 - \ln\frac{Q^2}{\mu^2} - 
\ln\left(\frac{1-\zeta}{\zeta}\right)\right) - 2 \ln\zeta \Big] \Bigg]. 
\end{align}

\end{document}